\documentclass{aa}

\usepackage{psfig}

\def\aa#1#2#3{#1,      {A\&A, }{\bf#2}, #3}
\def\apj#1#2#3{#1,     {ApJ, }{\bf#2}, #3}
\def\apjsupp#1#2#3{#1, {ApJS, }{\bf#2}, #3}
\def\mn#1#2#3{#1,      {MNRAS, }{\bf#2}, #3}

\begin{document}

\title{On the ionisation fraction in protoplanetary disks\\
III. The effect of X--ray flares on gas-phase chemistry}

\author{Martin~Ilgner and Richard~P.~Nelson}


\institute{Astronomy Unit, Queen Mary, Mile End Road, London E1 4NS, U.K. \\
\email{M.Ilgner@qmul.ac.uk, R.P.Nelson@qmul.ac.uk}}

\date{Received 29 March 2006 / Accepted 11 May 2006}

\titlerunning{Ionisation fraction in disks}

\authorrunning{M.~Ilgner, R.P.~Nelson}

\abstract
{ 
Recent observations of the X--ray emission from T Tauri
stars in the Orion nebula have shown that they undergo frequent outbursts
in their X--ray luminosity. These X--ray flares are characterised
by increases in luminosity by two orders of magnitude, a typical
duration of less than one day, and
a significant hardening of the X--ray spectrum.}
{It is unknown what effect these X--ray flares will have on the
ionisation fraction and dead--zone structure in protoplanetary disks.
We present the results of calculations designed to address this question.}
{We have performed calculations of the ionisation fraction in
a standard $\alpha$--disk model using two different chemical reaction
networks. We include in our models ionisation due to X--rays from the
central star, and calculate the time--dependent ionisation fraction 
and dead--zone structure for the inner 10 AU of a protoplanetary disk model.}
{We find that the disk response to X--ray flares 
depends on whether the plasma temperature increases during flares
and/or whether heavy metals (such as magnesium) are present in the gas phase.
Under favourable conditions the outer disk dead--zone can disappear altogether,
and the dead--zone located between $0.5 < R < 2$ AU can disappear and reappear
in phase with the X--ray luminosity.}
{X--ray flares can have a significant effect on the dead--zone
structure in protoplanetary disks. Caution is required in interpreting this 
result as the duration of X--ray bursts is considerably 
shorter than the growth time
of MHD turbulence due to the magnetorotational instability.}



\maketitle

\section{Introduction}
Observations of young stars in a variety of star forming regions,
ranging from the trapezium cluster in Orion to the Taurus--Auriga
complex, have shown that protostellar disks are ubiquitous 
(e.g. Beckwith \& Sargeant 1996; O'Dell et al. 1993; Prosser et al. 1994).
These disks often show evidence for active accretion with a canonical
mass flow rate onto the central star of $\sim 10^{-8}$ M$_{\odot}$ yr$^{-1}$ 
(e.g. Sicilia--Aguilar et al. 2004), requiring a mechanism to transport
angular momentum within the disks.
So far only one mechanism has been shown to work:
MHD turbulence generated by the magnetorotational instability (MRI)
(Balbus \& Hawley 1991; Hawley \& Balbus 1991).\\
\indent
Given that protostellar disks are cool and dense near their midplanes,
there are questions about the global applicability of the MRI to these
disks, as the ionisation fraction is expected to be low (Blaes \& Balbus 1994;
Gammie 1996). Indeed, nonlinear magnetohydrodynamic simulations of disks
that include ohmic resistivity (Fleming, Stone \& Hawley 2000)
have indicated that for magnetic Reynolds numbers $Re_{\rm m}$ below a
critical value $Re_{\rm m}^{\rm crit}$, turbulence is not sustained and
the disks return to a near--laminar state whose internal stresses are
too small to explain the observed mass accretion rates onto T Tauri stars.\\[1em]
\indent
There have been a number of studies of the ionisation fraction in
protostellar disks. Gammie (1996) first suggested that disks may
have magnetically ``active zones'' sustained by thermal or cosmic ray ionisation,
adjoining regions that are ``dead--zones'' where the ionisation fraction is
too small to sustain MHD turbulence. Sano et al. (2000) examined this
issue using a more complex chemical model that included dust grains.
Glassgold et al. (1997) and Igea et al. (1999) examined the
role of X--rays as a source of ionisation in protostellar disks,
and highlighted doubts about whether Galactic cosmic rays
could penetrate into the inner regions of protostellar disks
because of the stellar wind. 
Fromang, Terquem \& Balbus (2002) examined the influence
of gas phase heavy metals and demonstrated the potential
importance of charge--transfer reactions, and
Semenov et al. (2004) studied disk chemistry and
the ionisation fraction using a complex reaction network drawn from the UMIST
data base.\\
\indent
Recent observations of the X--ray emission from T Tauri stars in the Orion
nebula
using the Chandra observatory (COUP - Chandra Orion Ultradeep Project)
have shown that in addition to
providing a characteristic X--ray luminosity at a level of $L_{\rm X} \sim 10^{30}$
erg s$^{-1}$,  young stars emit X--ray flares whose luminosity is 
$\sim 100$ times this value (e.g. Wolk et al. 2005;
Favata et al. 2005). These flares typically last for less than a day,
and are characterised by a sharp linear rise in luminosity,
followed by an exponential decay. The typical recurrence time is
about one week, and associated with the flares is a hardening of
the X--ray spectrum indicating a rise in the plasma temperature in the 
stellar corona from $k_{\rm B}T \simeq 3$ keV to typical values of 
$\simeq 7$ keV. In this paper we address the question of what effect 
these X--ray flares have on the ionisation fraction and structure
of dead--zones in protostellar disks.\\[1em]
\indent
In a recent paper (Ilgner \& Nelson 2006a) we compared the predictions
made by a number of chemical reaction networks about the
structure of dead--zones in standard $\alpha$--disk models.
This study included an examination of the reaction scheme
proposed by Oppenheimer \& Dalgarno (1974), and more complex
schemes drawn from the UMIST data base (Le Teuff et al. 1996).
In a follow-up paper (Ilgner \& Nelson 2006b) we examined the role
of turbulent mixing in determining the structure of dead--zones
in $\alpha$--disk models using chemical reaction networks
drawn from (Ilgner \& Nelson 2006a). In this paper we continue
with our work on the ionisation structure within protoplanetary disks
and examine the 
effect that X--ray flares have on dead--zones in $\alpha$--disk models
using reaction networks drawn from Ilgner \& Nelson (2006a).\\
\indent
In general we find that X--ray flares can have a fairly dramatic
effect on the ionisation structure in disks, especially if the 
X--ray spectrum hardens during flares and/or trace quantities of heavy
metals (magnesium) are present in the gas phase. Our disk models
can be divided into three distinct regions: an inner region where
the disk is always active due to thermal ionisation; a central region
in which the dead--zone formally decreases in depth substantially or
disappears altogether
during X--ray flares, but which returns to being a deep dead--zone 
in between flares; an outer region beyond $R=2$ AU in which
the dead--zone depth does not change in time, and which can become very thin
or disappear altogether in the presence of heavy metals and
an increasing plasma temperature during outbursts.\\[1em]
\indent
This paper is organised as follows.
In section~\ref{sec2} we describe our modelling procedure, including the disk
model, the chemical models, and our method for simulating
X--ray flares. In section~\ref{sec3} we present the results of our
models, and discuss the effects of changing the model parameters.
In section~\ref{sec4} we discuss our results in the context of
turbulent protostellar disks, and in section~\ref{sec5} we
summarise our findings.

\section{Model}
\label{sec2}

\subsection{Disk model}
The underlying disk model considered is a standard $\alpha$--disk.
Details are given in Ilgner \& Nelson (2006a) and references therein. 
To recap: the disk is assumed to orbit a young solar mass star and 
undergo viscous evolution. We use the $\alpha$ prescription for the 
viscous stress, such that the kinematic viscosity 
$\nu= \alpha c_s^2/\Omega$, where $c_s^{}$ is the sound speed and 
$\Omega$ is the local Keplerian angular velocity. Heating of the disk 
is provided by viscous dissipation alone, and cooling by radiation 
transport in the vertical direction. The disk structure is obtained 
by solving for hydrostatic and thermal equilibrium. The disk model is 
completely specified by the mass accretion rate, ${\dot M}$ and the 
value of $\alpha$. In this paper we consider a single disk model with 
${\dot M}=10_{}^{-7}$ M$_{\odot}$ yr$_{}^{-1}$ and $\alpha=10_{}^{-2}$. 
The mass is 0.0087 M$_{\odot}$ between  $0.1 \le R \le 10 \rm \ AU$. 

\subsection{Kinetic models}
\label{sec:kinetic}
We have applied two kinetic models to evolve the gas--phase chemistry. 
Both reaction networks have been described in Ilgner \& Nelson (2006a), 
where the models were given the labels \texttt{model1} and \texttt{model3}, 
respectively. For continuity, we maintain this labelling convention in this
paper. To recap: \texttt{model1} refers to the kinetic model of 
Oppenheimer \& Dalgarno (1974) while \texttt{model3} is linked to the 
more complex UMIST database.
The underlying kinetic scheme of the Oppenheimer \&
Dalgarno model, involves two elements, five species, and four reactions.
These species are free electrons ``$e^-_{}$", a representative molecule ``m", a heavy
metal atom ``M", and their ionized counterparts ``m$^+_{}$" and ``M$^+_{}$". We used
the reference values of the rate coefficients given in Ilgner \& Nelson (2006a).
The UMIST kinetic model was constructed by
extracting all species and reactions from a set of 174 species in
the UMIST database containing the elements H, He, C, O, N, S, Si, Mg, Fe,
resulting in 1965 reactions being included.
A detailed description of both kinetic models
is given in Ilgner \& Nelson (2006a).\\
Apart from the ionisation rate $\zeta$ which 
is discussed in a separate section below, all the other parameters are taken 
from Ilgner \& Nelson (2006a).

\subsection{X--ray flares}
\label{ionisation}
We assume that ionisation of the disk material arises because of 
incident X--rays that originate in the corona of the central T 
Tauri star. We neglect contributions from Galactic cosmic rays 
as it remains uncertain whether they can penetrate into inner disk regions 
we consider due to the stellar wind. 
In our previous work (Ilgner \& Nelson 2006a, 2006b) we
introduced the X-ray luminosity $L_{\rm X}^{}$, which was constant.
In this work the X-ray luminosity becomes time dependent, $L_{\rm X}^{}(t)$
because of the X--ray flares.

\begin{figure}[t]
   \psfig{silent=,figure=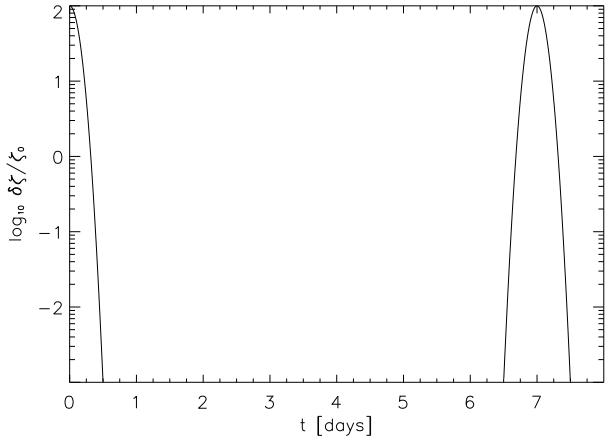,width=9.0cm}
   \caption[]{Time sequence of the X--ray flares $\delta \zeta/\zeta_0$.
 The perturbation is periodic in time with a
periodicity $\tau_{\rm X}^{} = 7 \ \rm days$ while the width
of a single perturbation is $t_{\sigma}^{} = 12 \rm \ hr$.
\label{figure1}}
\end{figure}

We now adopt a model in which the 
X--ray luminosity maintains a base value, $L_{\rm X}^{} = L_{\rm X}^{0}$,
on top of which are superposed X--ray flares with peak luminosity
$L_{\rm X}^{P}$.
In order to maintain the compatibility with 
our previous work (Ilgner \& Nelson 2006a, 2006b) we use 
$L_{\rm X}^{0} = 10_{}^{30} \ \rm erg/s$. 
\indent
We approximate the flare temporal morphology by a sequence of 
outbursts that arise periodically in time. The shape of a single flare is 
modelled by using a Gaussian profile. We note that the 
symmetric profile thus obtained is simpler than the observed flare morphology
which is more accurately 
characterised as a linear rise and exponential decay, but we believe
that our model captures the essentials of how X--ray flares affect
the ionisation structure in the disk independently of such details. 
We assume that one flare occurs per week, and each outburst lasts for 24 hours. 
These values are similar to those observed during the Chandra Orion Ultradeep
Project (COUP) as reported by Wolk et al. (2005).
\\
\indent
For a given total X--ray luminosity $L_{\rm X}^{}$ and plasma
temperature $k_{\rm B}^{}T$ one can calculate the ionisation rate 
due to X--rays at each position in the disk model. 
This requires an integraton along each line of sight through
the disk model to the X--ray source, and our method for this is
described in Ilgner \& Nelson (2006a).
In this paper we consider both models in which the plasma temperature
remains constant, and models in which the plasma temperature varies
along with the X--ray luminosity. 
When the plasma temperature remains constant the attenuation of
the X--rays is also constant, so the temporal morphology of the
X--ray luminosity and the local ionisation rate are 
characterised by the same mathematical function.  
The perturbation 
$\delta \zeta$ in the local ionisation rate due to an X--ray flare is 
then given by the following time-dependent ionisation rate
\begin{equation}
\zeta(t) = \zeta_0^{} + \delta \zeta(t),
\label{equation:1}
\end{equation}
where
\begin{equation}
\delta \zeta(t) = \left \{
\begin{array}{ll}
\lambda \zeta_0 \exp \left\{ - \left( \frac{t - \Delta}{\sigma} \right)_{}^2 \right\} &
\mathrm{if} \ 
\Delta - t_{\sigma}^{} \le t \le 
\Delta + t_{\sigma}^{} \\
0 & \rm otherwise
\end{array}
\right.
\label{equation:2}
\end{equation} 
where $\zeta_0^{}$ and $\delta \zeta$ denotes the 
base X--ray ionisation rate and its time-dependent 
perturbation, respectively. $\lambda$ is a factor that determines
the amplitude of the perturbation, 
$t_{\sigma}^{} =  12 \ \rm hr$, 
$\Delta = m \tau_{\rm X}^{}$ with $m$ as an integer, and 
$\tau_{\rm X}^{} = 7 \ \rm days$. Hence the characteristic
width $t_{\sigma}^{}$ of a single flare is given by 
\begin{equation}
\sigma = \frac{t_{\sigma}}{ \sqrt{|\ln [\delta \zeta(t_{\sigma}^{})/\lambda]|}}
\label{equation:3}
\end{equation}
with $\delta \zeta/\lambda = 10_{}^{-5}$ for $|t -\Delta |= t_{\sigma}^{}$. A time 
sequence of the modulation of $\delta\zeta$ with $\lambda = 100$ is shown 
in figure~\ref{figure1}.

Observations indicate that the plasma temperature $k_{\rm B} T$ increases
as the X--ray luminosity does during a flare. This means that
we need to take account of the fact that the penetration depth
also varies with time, due to the hardening of the X--ray spectrum.
As we will see in section~\ref{sec3.3}, this can
have quite dramatic effects on the local ionisation rate.
To model this we consider a minimum plasma temperature $T_{\rm cool}$
which applies when the X--ray luminosity is at its base value
$L_{\rm X}^0$ and calculate the ionisation rate $\zeta_{\rm 0}(R,z)$
at each position in the disk. When the X--ray luminosity has reached its
peak value $L_{\rm X}^P$ we assume that the plasma temperature has reached
its maximum value $T_{\rm hot}$, and calculate the ionisation
rate $\zeta_{\rm P}(R,z)$ at each position in the disk.
The time dependent perturbation to the local ionisation rate is then
given by
\begin{equation}
\delta \zeta(t)=\zeta_{\rm P}
\exp \left\{ - \left( \frac{t-\Delta}{\sigma} \right)_{}^2 \right\}
\label{equation:4}
\end{equation}
if $| t - \Delta | < t_{\sigma}$, otherwise $\delta \zeta=0$.
In the models with varying plasma temperature, we adopt
$k_{\rm B}T_{\rm cool}=3$ keV
and $k_{\rm B} T_{\rm hot}=7$ keV, which are close to the
observed values (Wolk et al. 2005; Favata et al. 2005)
\begin{figure*}[t]%
\hbox{%
\psfig{silent=,figure=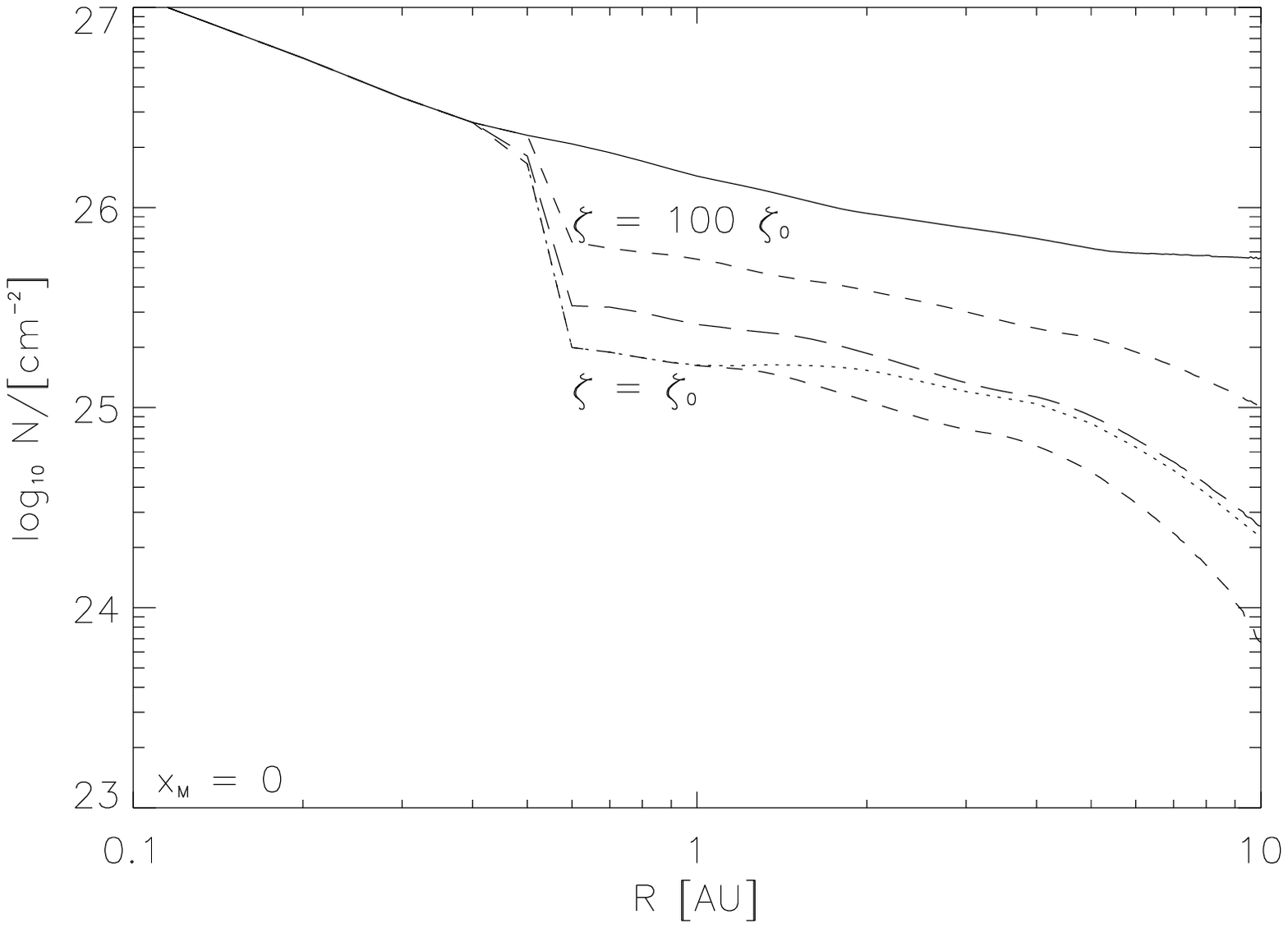,width=9.0cm} \hspace*{-.6cm}
\psfig{silent=,figure=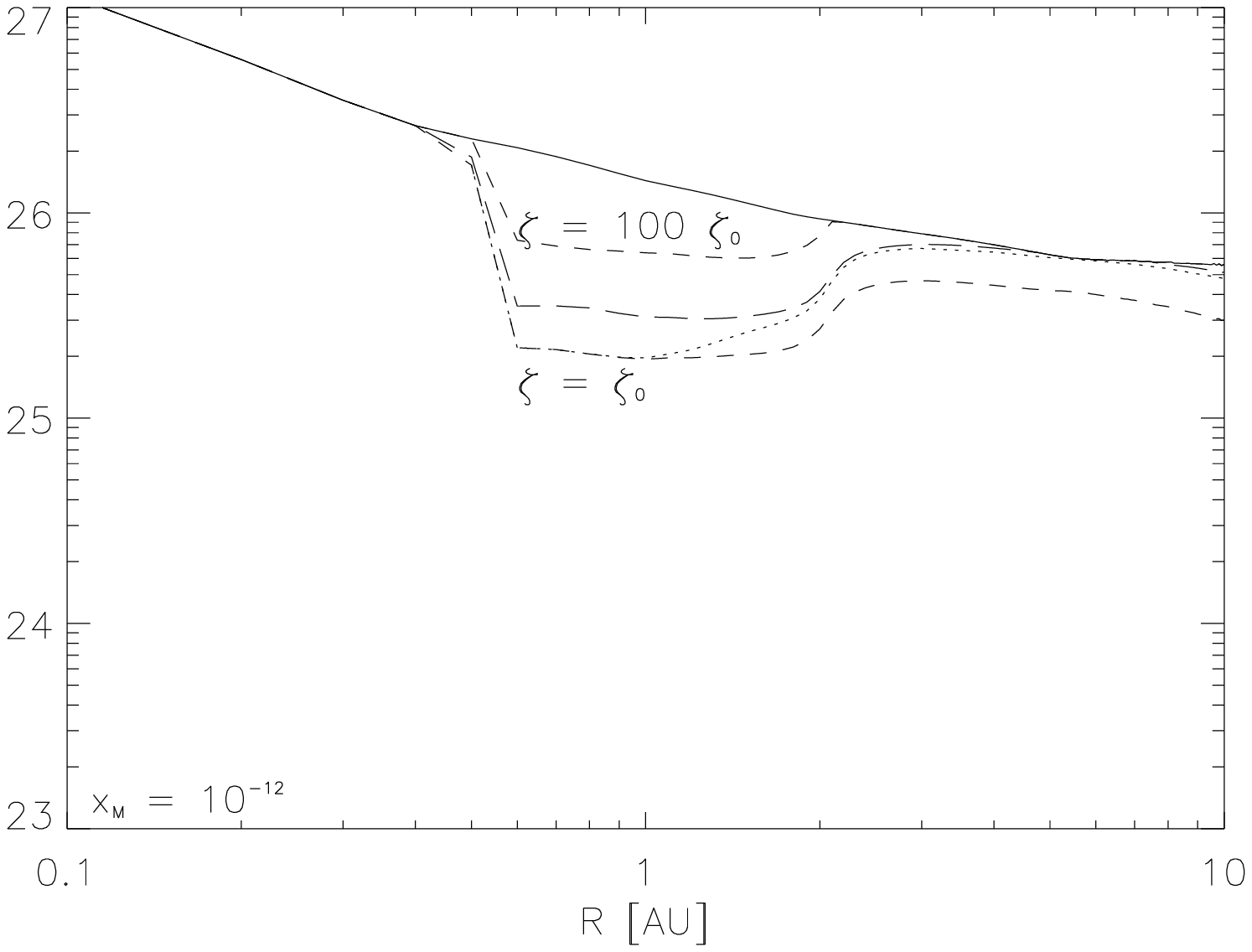,width=9.0cm}} \vglue0.0cm
\caption{\texttt{model1} - Column densities of the whole disk (solid line)
and of the active zones (dashed and dotted lines) - corresponding to magnetic
Reynolds numbers $Re_{\rm m}$ greater than 100 - for two different heavy metal
abundances $x_{\mathrm{M}^{}}$. The left panel is for $x_{\rm M}=0$
and the right panels is for $x_{\rm M}=10^{-12}$.
The upper and lower dashed lines refer
to simulations with constant ionisation rate $\zeta = \zeta_0$
and $\zeta = 100 \times \zeta_0$ while the long-dashed line refers to
simulation with $\zeta = \zeta_{\rm eff}$.
The dotted line refers to the column density
of the active zone at $t = 10, 000 \ \rm yr$ obtained by assuming a
time-dependent
ionisation rate $\zeta(t) = \zeta_0^{} + \delta \zeta(t)$. The time of the
plot is at the mid-point between two X--ray flares.
\label{figure2}}
\end{figure*}

\section{Results}
\label{sec3}
We assume $L_{\rm X}^{0} = 10^{30}_{} \ \rm erg/s$ for the 
base value of the X--ray luminosity, in basic agreement with
recent observations. During the X--ray flare the 
X--ray peak luminosity $L_{\rm X}^{P}$ is significantly larger 
then $L_{\rm X}^{0}$. We assumed $L_{\rm X}^{P} = 100 \times 
L_{\rm X}^{0}$ which is in line with the observed changes reported 
in Fatava et al. (2005). 
The ionisation rate $\zeta$ is then given by equations~(\ref{equation:1}) 
and~(\ref{equation:2}) with $\lambda = 100$ if $k_{\rm B} T$ is constant,
or equations~(\ref{equation:1}) and (\ref{equation:4}) if $k_{\rm B} T$
varies. The time averaged value over one
complete seven day cycle is 
$< \! \zeta \!> \ = \zeta_{\rm eff} = \ 5.81 \times \zeta_0^{}$ when
$k_{\rm B} T$ has a constant value of 3 keV.\\
\indent
We have evolved the disk chemistry using the kinetic models 
\texttt{model1} and \texttt{model3} by taking the perturbation $\delta \zeta$ 
of the ionisation rate due to X--ray flares into account. As in our previous 
studies we wish to determine which parts of the disk are sufficiently 
ionised for the gas to be well coupled to the magnetic field, and thus 
able to maintain MHD turbulence, and which regions are too neutral 
for such turbulence to be maintained. Again, we refer to those regions 
as being ``active'' and ``dead'' zones respectively, with the region 
bordering the two being the ``transition'' zone. The important 
discriminant that determines whether the disk is active or dead is 
the magnetic Reynolds number, ${Re}_{\rm m}^{}$, defined by
\begin{equation}
{Re}_{\rm m}^{} = \frac{H c_s}{\mu}
\label{reynolds}
\end{equation}
where $H$ is the disk semi--thickness, $c_s$ is the sound speed, 
and $\mu$ is the magnetic diffusivity which is a function of the local
free electron fraction. We adopt a value of 
${Re}_{\rm m}^{\rm crit} = 100$ in this paper, following the value 
used in our previous publications and Fromang et al. (2002). We are 
able to calculate the distribution of ${Re}_{\rm m}^{}$ within our 
disks. Regions with ${Re}_{\rm m}^{} < 100$ are deemed to be 
magnetically dead, and those with ${Re}_{\rm m}^{} > 100$ magnetically 
active. The quantity $x_{\rm crit}^{}[\rm e_{}^{-}]$ denotes the ionisation 
fraction $x[\rm e_{}^{-}]$ along the transition zone.\\
\indent
Our approach to modelling the chemistry in disks with X--ray flares
is as follows.
First, we evolved models
keeping the ionisation rate $\zeta$ constant in time.
We considered three cases with 
$\zeta = \zeta_0$, $\zeta=\zeta_{\rm eff}^{}$, and 
$\zeta = 100 \times \zeta_0$ 
The kinetic equations are solved for a time 
interval of $\Delta t = 100, 000 \rm \ yr$. Hence, the ionisation 
fraction $x[\rm e_{}^-]$ is a function of time $t$, and in principle 
so is the location of the transition zone. However, for these models which 
assume a time-independent ionisation rate $\zeta$, the change in the 
vertical location of the transition zone at all cylindrical radii in 
the computational domain was below the grid resolution for 
$t > 10, 000 \ \rm yr$.\\
\indent
Using the abundances obtained from the $\zeta = \zeta_0$ model at
$t = 100, 000 \rm \ yr$ as initial abundances,
we also calculated the ionisation fraction $x[\rm e_{}^-]$
by considering a time-dependent ionisation rate
$\zeta = \zeta_0 + \delta \zeta $.
The kinetic equations here were
solved for a time interval of $\Delta t = 10, 000 \rm \ yr$. We restricted 
the time integration of the kinetic equations by assuming a finite maximum 
absolute step size $\Delta h_{\rm max}^{} = 1 \ \rm hr$ in order to resolve 
the flares in time. 
Because of the adopted value of the perturbation $\lambda=100$,
models with time-independent 
ionisation rates $\zeta = \zeta_0^{}$ and 
$\zeta = 100 \times \zeta_0^{}$ represent the limiting cases for models 
with $\zeta = \zeta_0^{}+ \delta \zeta$.

\subsection{Snapshots at $t = 10,000 \rm \ yr$ with $k_{\rm B} T$ constant}
\label{sec3.1}
We begin by discussing models which have a constant plasma temperature
$k_{\rm B} T=3$ keV but varying X--ray luminosity. 
The effects of allowing the plasma temperature to
vary are discussed in section~\ref{sec3.3}.
The discussion below relates to the disk properties at a single point in time
midway between two X--ray flares. Discussion about the time dependent behaviour
of the ionisation fraction and dead--zones is presented in section~\ref{sec3.2}.\\
\begin{figure*}[t]%
\hbox{%
\psfig{silent=,figure=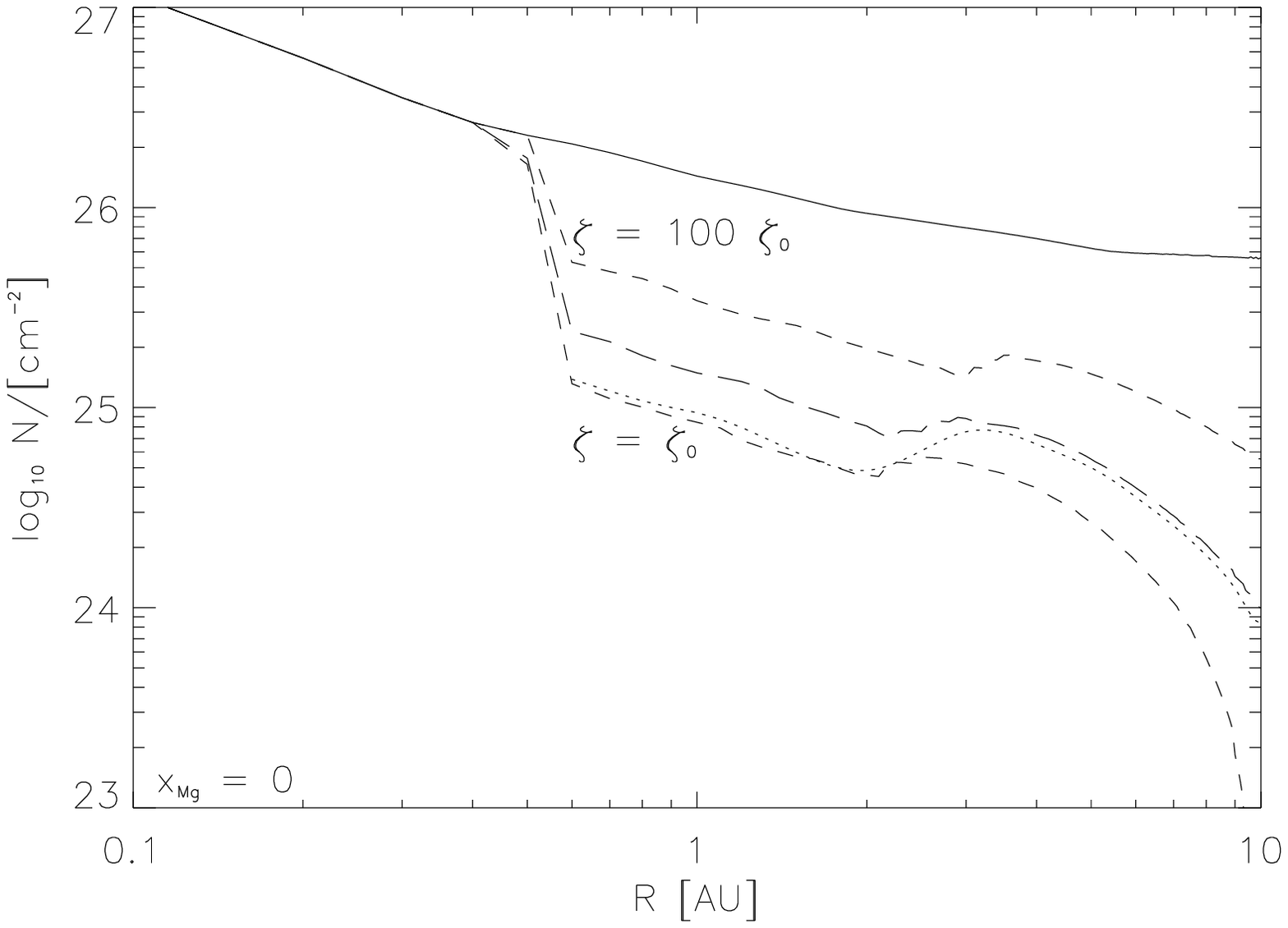,width=9.0cm} \hspace*{-.6cm}
\psfig{silent=,figure=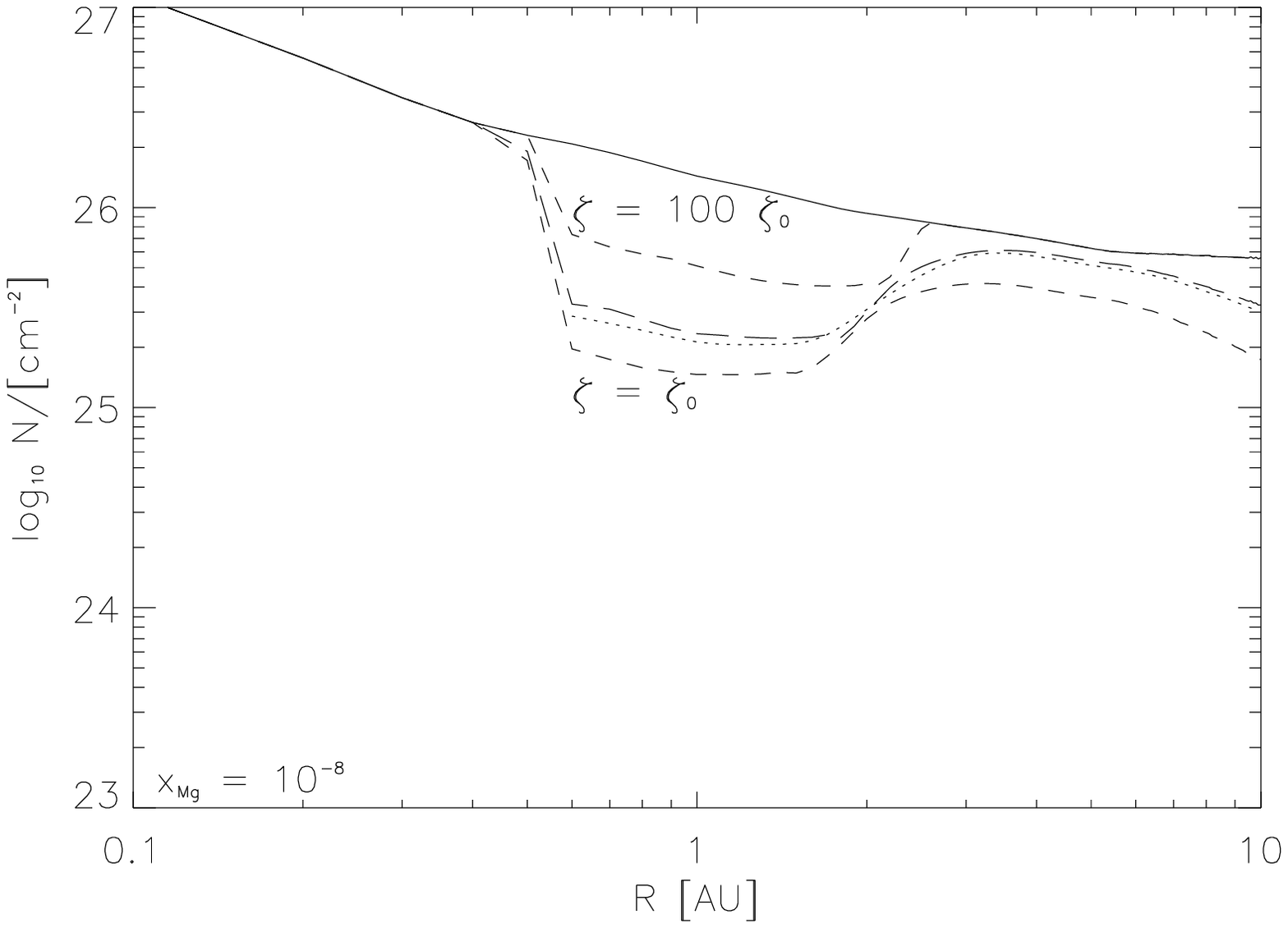,width=9.0cm}} 
\caption{\texttt{model3} - Column densities of the whole disk (solid line) 
and of the active zones (dashed and dotted lines) - refering to magnetic 
Reynolds numbers $Re_{\rm m}$ greater than 100 - for two different elemental 
abundances $x_{\mathrm{Mg}^{}}$.
The left panel is for $x_{\rm Mg}=0$ and the right panel is for 
$x_{\rm Mg}=10^{-8}$. The upper and the lower the dashed lines refer 
to simulations assuming a constant ionisation rate $\zeta = \zeta_0$ 
and $\zeta = 100 \times \zeta_0$, while the long-dashed dotted line 
refers to simulation with $\zeta = \zeta_{\rm eff}$. The dotted line shows 
the column density of the active zone at $t = 10, 000 \ \rm yr$ assuming a 
time-dependent ionisation rate $\zeta(t) = \zeta_0^{} + \delta \zeta(t)$. 
The amplitude $\lambda$ of the perturbation $\delta \zeta/\zeta_0$ is $\lambda = 100$ caused 
by increasing the X--ray luminosity 
$L_{\rm X}^{P} = 100 \times L_{\rm X}^{0}$ 
at the flare peak. 
\label{figure3}}
\end{figure*}

\noindent
{\bf Oppenheimer \& Dalgarno model}

\noindent
The results obtained for \texttt{model1} are presented in 
fig.~\ref{figure2}, 
which shows the column density of the whole disk plotted as a
function of radius 
using the solid line, and the column density of the active zone using either 
dashed lines (for which 
$\zeta = \zeta_0$, $\zeta = \zeta_{\rm eff}^{}$, and
$\zeta = 100 \times \zeta_0$)
or dotted lines referring to 
$\zeta(t) = \zeta_0 + \delta \zeta(t)$. The dotted line corresponds to a time
which is halfway between two X--ray flares.
The left panel shows cases for which the heavy metal abundance
$x_{\rm M}=0$ and the right panel shows cases with $x_{\rm M}=10^{-12}$.\\
\indent
As expected, figure~\ref{figure2} shows that a time-independent increase in 
the X--ray luminosity leads to a corresponding decrease in the depth of
the dead--zone (note that the disk inner regions are fully active because
of thermal ionisation of potassium [Ilgner \& Nelson 2006a]).
Interestingly, even an increase in X--ray luminosity by a factor of 100
is insufficient to fully ionise the disk, and a dead--zone remains beyond
$R> 0.5$ AU. When we consider the X--ray flaring model, we find that the
behaviour of the ionisation fraction and dead--zone depends on radial
position within the disk. The region between $0.5 < R < 1.2$ AU has the same
dead--zone structure as the model with constant X--ray luminosity set at the 
base level $L_{\rm X}^0$. As will be discussed in more detail later 
on, this is because the recombination time in this region is shorter than the 
period of the X--ray outburst cycle, so the disk ionisation fraction remains 
in phase with the X--ray luminosity. Further out in the disk beyond $R > 2$ 
AU, we find that the dead--zone structure is very close to that obtained when 
the X--ray luminosity takes a constant value that is equal to the time--averged 
value of the X--ray flaring model (i.e. $\zeta=\zeta_{\rm eff}$). This is 
because the recombination time is now longer than the period of the X--ray  
flaring, so the response of the ionisation fraction lags the instantaneous 
ionisation rate. Over long evolution times the disk responds to the average 
ionisation rate.\\
\indent
The right-hand panel in figure~\ref{figure2} shows the effect of introducing
a small abundance ($x_{\rm M}=10^{-12}$) of heavy metals. As described in
Fromang et al. (2002) and Ilgner \& Nelson (2006a), the introduction of 
heavy metals to the gas phase
is expected to increase the free--electron fraction
because of charge--transfer reactions with molecular ions.
Figure~\ref{figure2} shows that an increase in constant X--ray luminosity
by a factor of 100 causes the dead--zone to disappear beyond 2 AU.
A constant ionisation rate $\zeta_{\rm eff}$ corresponding to the time-average
of the flaring rate leads to effective removal of the dead--zone beyond
3 AU. When we consider the X--ray flaring model we observe similar behaviour
to that in the model without heavy metals. Interior to about 1 AU
the dead--zone structure at the time of the snapshot is essentially the
same as the one obtained when the ionisation rate has a constant value
$\zeta = \zeta_0$. As discussed in Ilgner \& Nelson (2006a), the recombination
of free electrons in this region remains dominated by molecular ions even when 
heavy metals are present, so this result is expected. Further out in the disk
the recombination becomes dominated by the heavy metal ions, M$^+$,
and the recombination time is longer than the period of the X--ray flares.
In this region the disk again responds to the time dependent X--ray flares
as if the ionisation rate were equal to the time averaged value, 
resulting in the effective removal of the dead--zone beyond $R \simeq 3$ AU. \\

\noindent
{\bf UMIST model}

\noindent
The results obtained for \texttt{model3} are presented in fig.~\ref{figure3}. 
Here, the column density of the whole disk is plotted as a function of radius 
\begin{figure*}[t]%
\hbox{%
\psfig{silent=,figure=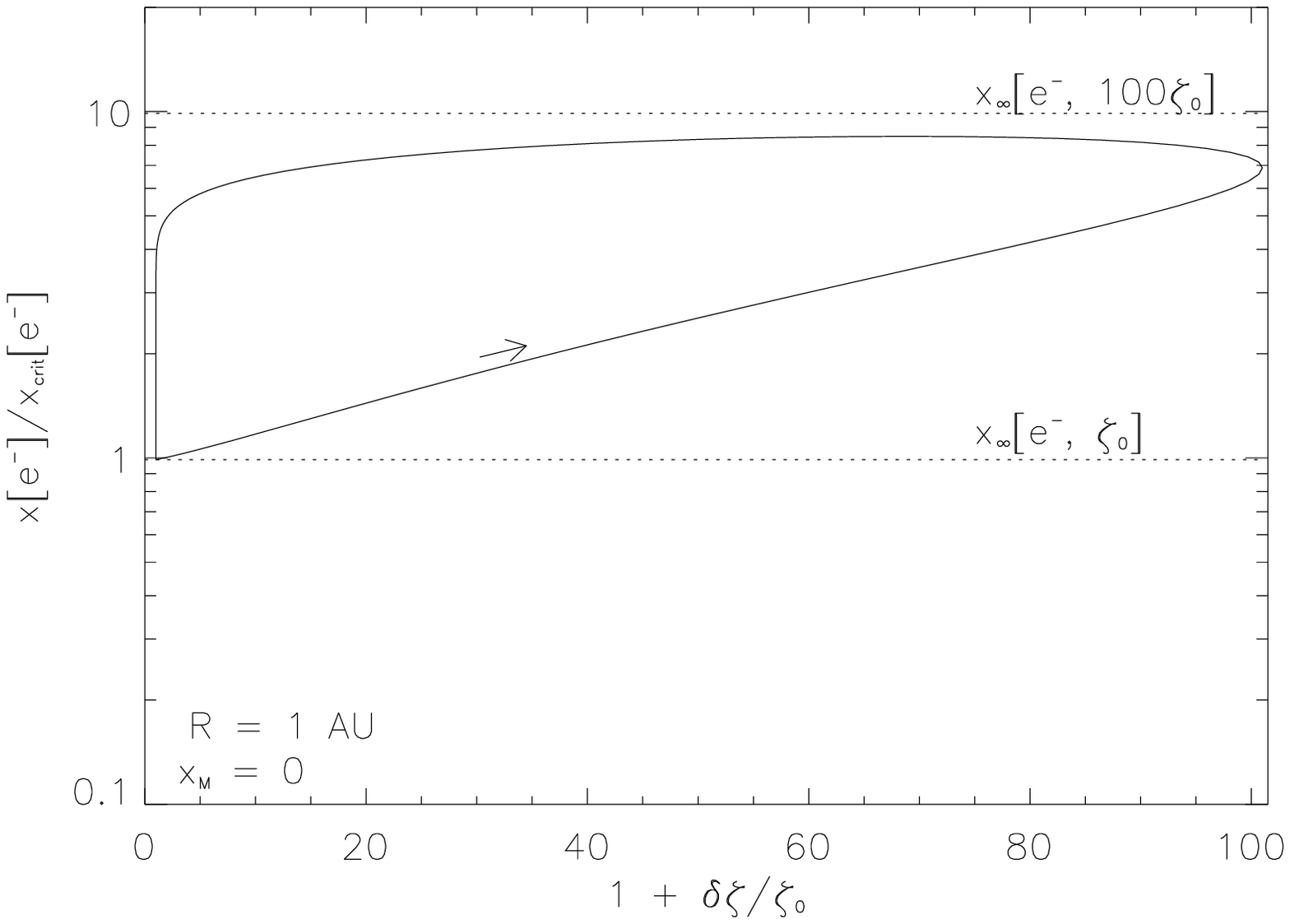,width=9.0cm} \hspace*{-.6cm}
\psfig{silent=,figure=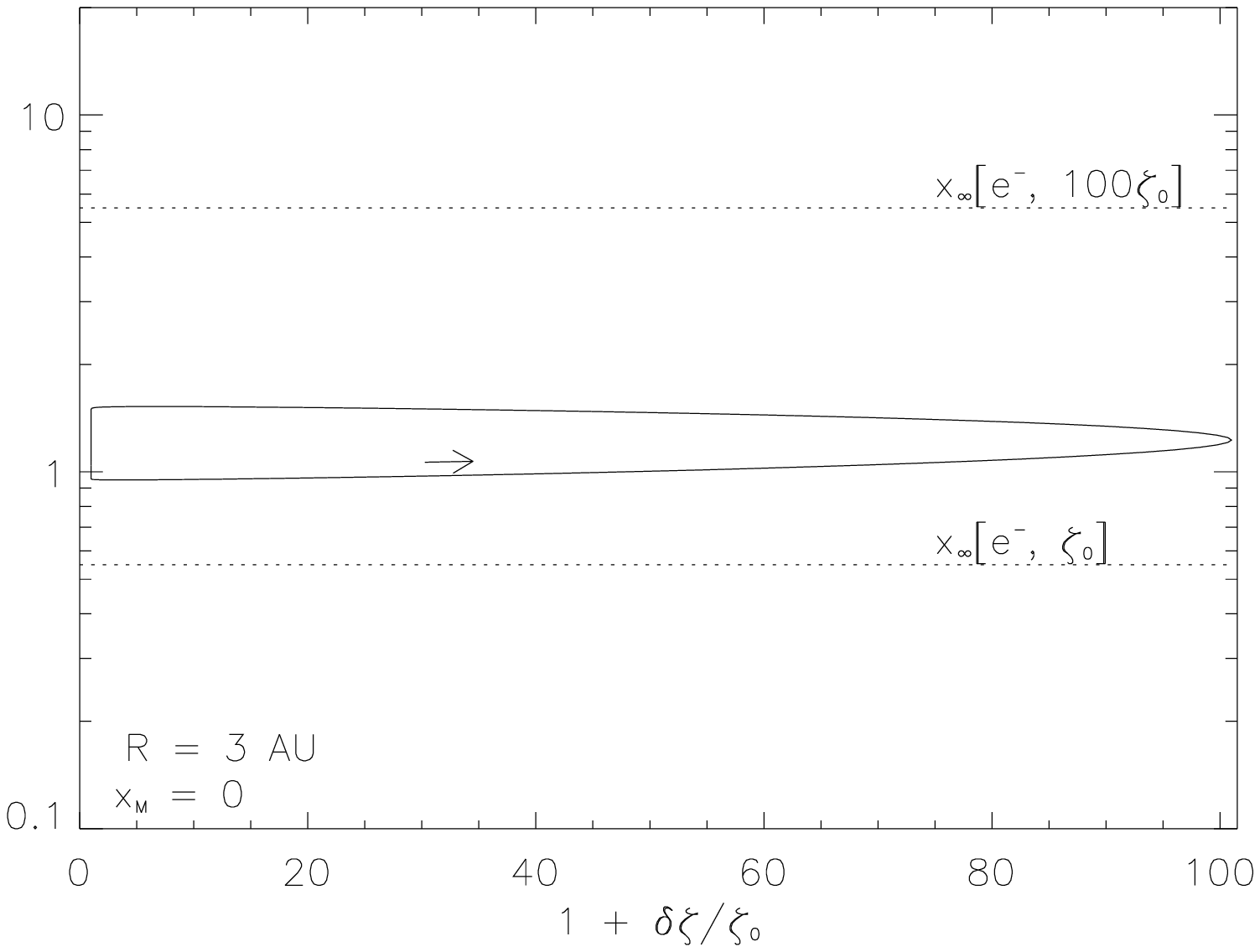,width=9.0cm}} 
\caption{\texttt{model1} - The change in the ionisation fraction 
$x[\rm e_{}^{-}]/x_{\rm crit}[ e_{}^{-}]$ is plotted against the relative change in 
the ionisation rate $(\zeta_0 + \delta \zeta)/\zeta_0$ at different 
radial positions $R$ along the 
transition zone assuming $x_{\rm M} = 0$. 
It is a limit cycle since the orbit is 
closed (the arrow points forward in time). 
In addition, the equlibrium values 
$x_{\infty}[\rm e_{}^{-}]$ are shown obtained for the unperturbed models with 
$\zeta = \zeta_0$, and $\zeta = 100 \times \zeta_0$.
\label{figure4}}
\end{figure*}
using the solid line while the dashed lines correspond to 
the column density of the active zone obtained for models with 
$\zeta = \zeta_0^{}$, $\zeta = \zeta_{\rm eff}$, and 
$\zeta = 100 \times \zeta_0^{}$. In addition, the column density of 
the active zone for the model employing the time dependent ionisation 
rate $\zeta = \zeta_0^{} + \delta \zeta$ is shown by the dotted line.
The left panel  corresponds to models in which the elemental 
abundance of magnesium $x_{\rm Mg}=0$, the right panel shows results
for which $x_{\rm Mg}=10^{-8}$. \\
\indent
Our previous work that compared the results of different chemical
networks has already established that the more complex chemical
models based on the UMIST data base predict 
deeper and more extensive dead--zones than the simpler Oppenheimer \& Dalgarno 
models (Ilgner \& Nelson 2006a). The primary reason is the larger number of
molecular ions that occur in the complex chemistry, leading to a faster
recombination of free electrons. Comparing figures~\ref{figure1} 
and \ref{figure3} confirms this. Considering the response
of \texttt{model3} to differing constant X--ray fluxes, we observe that
very similar trends arise as in \texttt{model1}. For example, an increase
in X--ray luminosity by a factor of 100 still leaves a substantial dead--zone
beyond $R> 0.5$ AU.\\[1em]
\indent
Considering \texttt{model3} with X--ray flares, we see that the dead--zone
between $0.5 < R < 2$ AU is of similar depth to that generated by
the base value of the X--ray luminosity $L_{\rm X}^0$
(with associated ionisation rate
$\zeta_0$). This arises for similar reasons to those given
when describing \texttt{model1} above: the local recombination time is shorter
than the time period between X--ray flares, ensuring that the instantaneous
ionisation fraction remains more or less in phase with the X--ray luminosity.
Conversely, the dead--zone depth beyond $R>2$ AU approaches that 
predicted by the model whose ionisation rate $\zeta_{\rm eff}$ is the time
average of the X--ray flaring model. This arises because the recombination time
here is longer than the time period between X--ray flares, ensuring that
the local ionisation fraction responds to the average X--ray luminosity rather
than the instantaneous value.\\
\indent
The right panel shows the effect of adding an elemental abundance of
magnesium $x_{\rm Mg}=10^{-8}$. For a constant X--ray luminosity
increased above the base value by a factor of 100, the dead--zone
disappears beyond $R>2$ AU. The dead--zone is predicted to be very shallow
in this region by the X--ray flaring model. In both these cases,
however, a significant dead--zone remains between $0.5 < R < 2$ AU
because the dominant sources of recombination here are molecular
ions rather than magnesium ions.

\subsection{Time--dependent results beyond $t = 10,000 \ \rm yr$ 
with $k_{\rm B} T$ constant }
\label{sec3.2}
In the previous section~\ref{sec3.1} our analysis focused on the
chemical state of the disk at a particular point in time
midway between two X--ray flares after 10,000 yr of evolution.
Since the ionisation rate $\zeta(t) = \zeta_0^{} + \delta \zeta(t)$ 
changes 
periodically in time, so does the size of the active zone.
We now focus on this periodic time dependence.
We consider the time--dependent ionisation fraction at a few well--defined
positions within the disk, along the transition zone
defined by figures~\ref{figure2} and \ref{figure3}.
We refer to the ionisation fraction at the transition zone
(defined where the magnetic Reynolds number $=100$) as the
critical value, and denote it by $x_{\rm crit}[\rm e_{}^{-}]$.
\begin{figure*}[t]%
\hbox{%
\psfig{silent=,figure=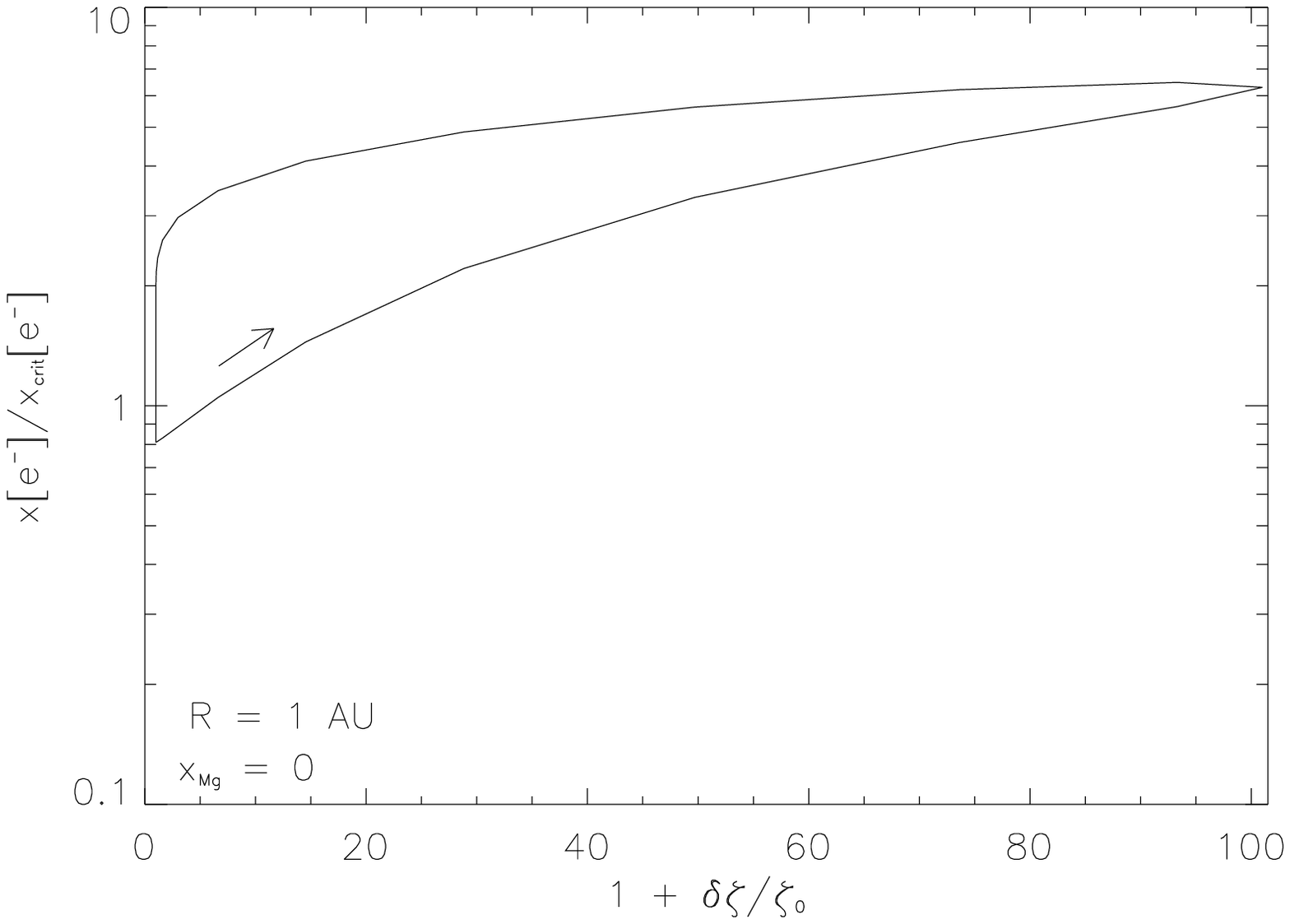,width=9.0cm} \hspace*{-.6cm}
\psfig{silent=,figure=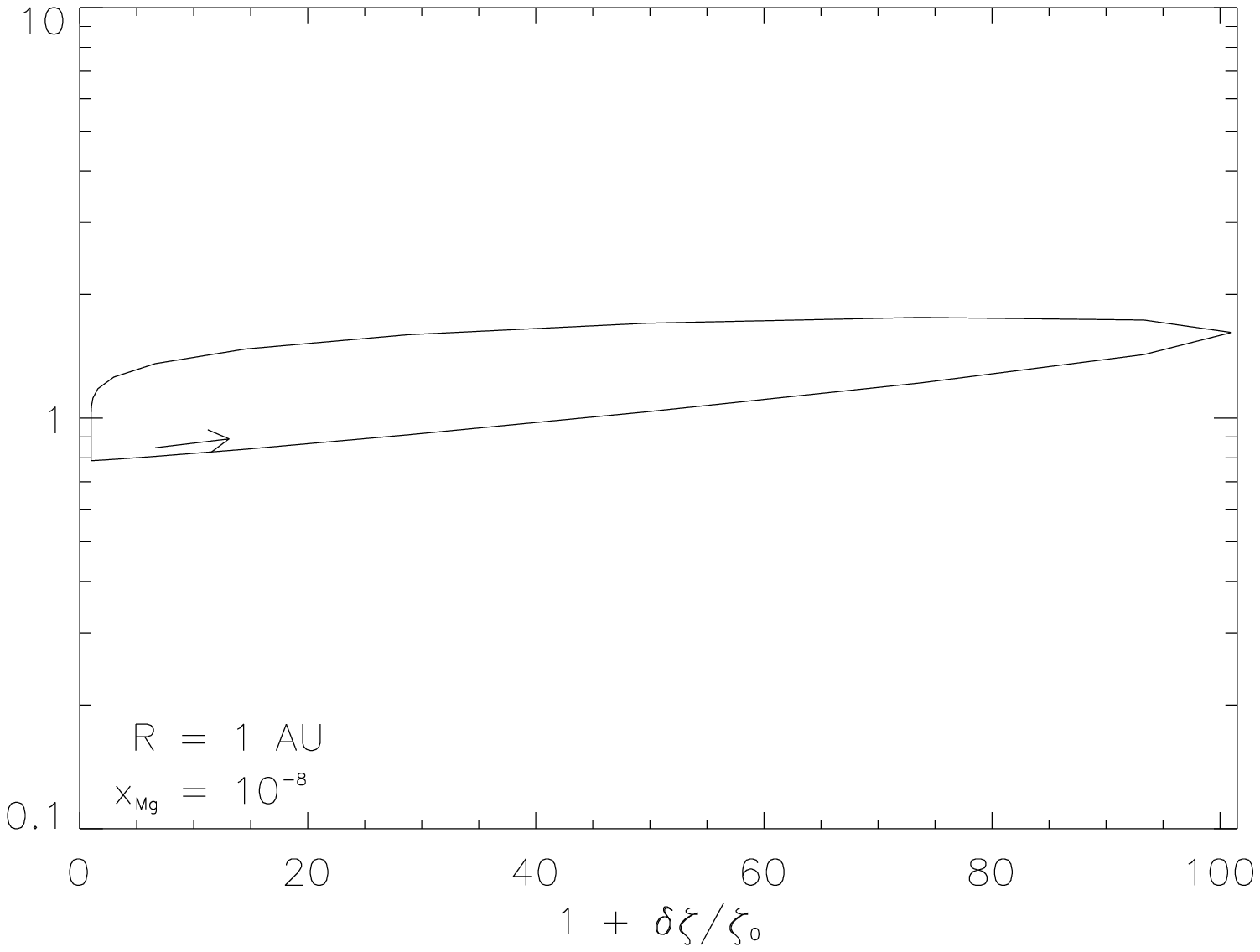,width=9.0cm}}
\caption{\texttt{model3} - The change in the ionisation fraction 
$x[\rm e_{}^{-}]/x_{\rm crit}[ e_{}^{-}]$ is plotted against the relative change in 
the ionisation rate $(\zeta_0+ \delta \zeta)/\zeta_0$ 
near the transition zone at $R = 1 \ \rm AU$ for 
different elemental abundances $x_{\rm Mg}$. The left panel is
for $x_{\rm Mg}=0$ and the right panel is for $x_{\rm Mg}=10^{-8}$.
It is a limit cycle since the orbit is 
closed (the arrow points forward in time). 
Note that $x[\rm e_{}^{-}]/x_{\rm crit}[ e_{}^{-}]$ is slightly less than 
unity when $\zeta=\zeta_0$ because the position of the transition zone
does not coincide precisely with a grid cell in this model.
\label{figure5}}
\end{figure*}
Again we discuss the results of the Oppenheimer \& Dalgarno model 
prior to the UMIST model.\\

\noindent
{\bf Oppenheimer \& Dalgarno model}\\
\noindent We found that the change in $x[\rm e_{}^{-}]$ with time $t$ is
given by a limit cycle. The limit cycle becomes apparent by plotting 
the ionisation fraction relative to the critical value  in the transition zone
$x[{\rm e}_{}^{-}]/x_{\rm crit}^{}[\rm e_{}^{-}]$ against the
change in the ionisation rate $(\zeta_0 + \delta \zeta)/\zeta_0$. 
The ionisation fraction repeats with a periodicity 
of $\tau_{\rm X}^{}$ resulting in a closed orbit. 
Examples of limit cycles obtained 
for \texttt{model1} in the absence of heavy metals
are shown in figure~\ref{figure4}; note 
that the arrow points 
forward in time such that the cycle is traced in an anticlockwise direction. 
The left hand panel shows the variation of $x[{\rm e}^-]$
at $R=1$ AU for \texttt{model1}.
The limit cycles obtained at $R \le 1 \ \rm AU$ with 
$x_{\rm M}^{} = 10_{}^{-12}$ are quite similiar to those obtained with 
$x_{\rm M}^{} = 0$ since the molecular ion still dominates the recombination of 
free electrons there. At locations where metal ions dominate when
they are included, such as at 3 AU (as shown in the right panel 
of figure~\ref{figure4}) the change in the ionisation
fraction $x[\rm e_{}^{-}]$ across the limit cycle is tiny and would appear as a 
straight line when using the same scale of figure~\ref{figure4}. 
The results of \texttt{model1} at 3 AU without heavy metals
shows a modest rise in ionisation fraction
during the cycle, as shown by the right panel of figure~\ref{figure4}.

The time required for the system 
to establish a limit cycle varies depending on the local position 
along the transition zone, but to within an accuracy of
$0.1 \%$ limit cycles were achieved throughout the disk for
\texttt{model1}, with and without heavy metals. \\
\indent
We differentiate between limit cycles for
which the change in ionisation fraction $x[\rm e_{}^{-}]$ lags substantially
behind the corresponding 
ionisation rate $\zeta(t)$, and those for which $x[\rm e_{}^{-}]$ does not. 
The former behaviour is illustrated
by the right panel of figure~\ref{figure4} where the value of
$x[\rm e_{}^{-}]$ continues to rise during the rise {\em and} fall of the
perturbed ionisation rate $(\zeta_0 + \delta \zeta)/\zeta_0$ 
during an X--ray flare. 
This phenomenon occurs only in those regions where the recombination
time is sufficiently long that the instantaneous recombination rate
is smaller than the instantaneous perturbed ionisation rate for the duration
of the flare. Once the flare has died away then the 
ionisation fraction slowly decreases back down to its original value,
just in time for the next flare to begin. This slow decrease in
$x[\rm e_{}^{-}]$ arises because the recombination time is
longer than the time period between X--ray flares.

The left panel of figure~\ref{figure4} 
shows an example where the changes in ionisation
fraction are more in phase with the perturbation to the ionisation rate.
This occurs in regions where the recombination time becomes short
compared to the perturbed ionisation rate during a flare.
The onset of an X--ray flare leads to
a quite dramatic rise in the ionisation fraction at the transition zone.
As the flare reaches its peak and begins to subside, the recombination
rate starts to exceed the ionisation rate and the ionisation fraction
decreases while the flare subsides. At the point where the flare ends
the perturbation to $x[\rm e_{}^{-}]$ has dropped from a peak value
of $x[{\rm e}_{}^{-}] \simeq 8 x_{\rm crit}[{\rm e}^{-}]$ down
to $x[{\rm e}_{}^{-}] \simeq 4 x_{\rm crit}[{\rm e}^{-}]$. Once the
flare has subsided completely the ionisation fraction drops
to the value corresponding to the steady state obtained when
the X--ray luminosity has its base value (with associated ionisation
rate $\zeta_0$). This final drop in the ionisation fraction
to the critical value $x[\rm e_{\rm crit}^{-}]$ at the transition zone occurs 
because the recombination time is shorter than the time period
between X--ray flares.

Figure~\ref{figure4} also shows that the duration of the
X--ray flares is too short to reach the ionisation fraction
obtained for \texttt{model1} with a constant ionisation rate
$\zeta=100 \zeta_0$.
Indeed we found that for the unperturbed models the transition from state 
$x_{\infty}[{\rm e}_{}^{-}; \zeta_0]$ to 
$x_{\infty}[{\rm e}_{}^{-}; 100 \times \zeta_0]$ occurs on time 
scales between $1 \ \rm day$ (at $R = 1 \ \rm AU$) and 
$90 \ \rm days$ (at $R = 10 \ \rm AU$), where $x_{\infty}$ denotes the
steady fractional abundance. \\[.5em]

\begin{figure*}[t]%
\hbox{%
\psfig{silent=,figure=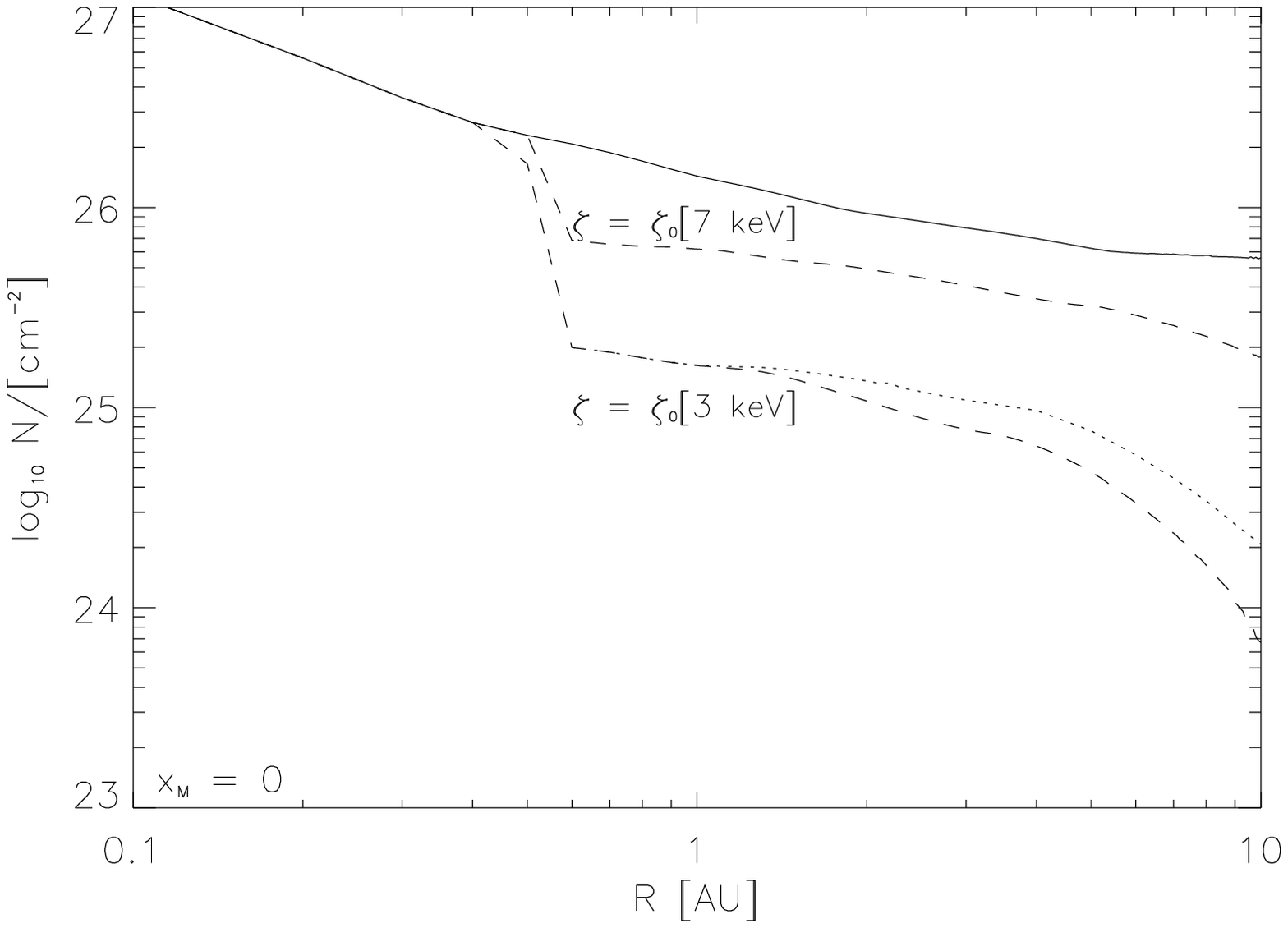,width=9.0cm} \hspace*{-.6cm}
\psfig{silent=,figure=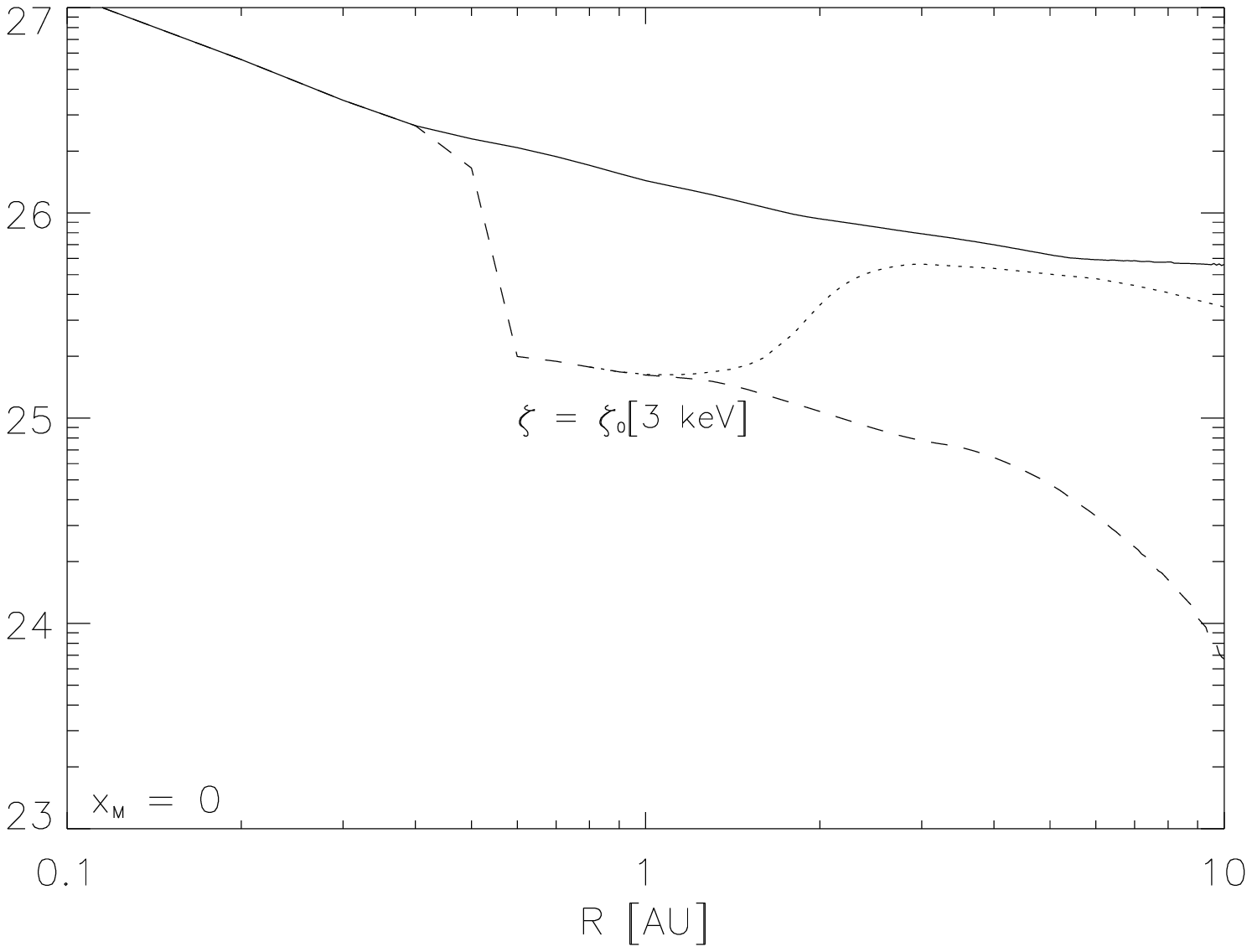,width=9.0cm}}
\caption{ - \texttt{model1} - Column densities of the whole disk (solid line)
and of the active zones (dashed and dotted lines) - refering to magnetic
Reynolds numbers $Re_{\rm m} > 100$ - for $x_{\rm{M}^{}} = 0$.
The dashed lines refer to simulations obtained by
assuming unchanging ionisation
rates $\zeta = \zeta_0$ for different plasma temperatures.\newline
\underline{left panel}: The dotted line refers to
the column density of the active
zone at $t = 10, 000 \ \rm yr$ obtained by assuming a time-dependent ionisation
rate caused by periodic variation of the plasma temperature between
$k_{\rm B} T_{\rm cool}=3$ keV and $k_{\rm B} T_{\rm hot}=7$ keV.\newline
\underline{right panel}: The dotted line refers to the column
density of the active
zone at $t = 10, 000 \ \rm yr$ obtained by assuming a time-dependent ionisation
rate $\zeta(t)$ caused by both the X--ray luminosity and the plasma temperature
increasing during flares.  The disk becomes entirely active when the
X-ray luminosity is increased by a factor of 100 and the plasma
temperature is maintained at a constant value of $k_{\rm B} T=7$ keV.
\label{figure6}}
\end{figure*}

\noindent {\bf UMIST model}

\noindent
We now continue by discussing the time dependent evolution of the 
ionisation fraction 
$x[\rm e_{}^{-}]$ obtained for \texttt{model3} at positions in the 
immediate vicinity of the transition zone for times beyond 10,000 yr.
Once again we found limit cycle behaviour throughout the disk, but in
order to obtain periodic orbits beyond radii $R> 2$ AU we had to evolve
the disk slightly beyond 10,000 yr (since a steady-state had not been reached
by 10,000 yr).
Figure~\ref{figure5} shows examples of the limit cycles
obtained at $R=1$ AU in the absence of magnesium (left panel) and
with an abundance of magnesium $x_{\rm Mg}=10^{-8}$ (right panel). 
For that region, metals do not have a dramatic effect 
on the ionisation fraction $x[\rm e_{}^{-}]$ because molecular ions 
still dominate the recombination process of free electrons (see Ilgner \& 
Nelson 2006a). It is clear that in both cases the free electron
fraction remains close to being in phase with the changing X--ray luminosity,
with $x[{\rm e}^{-}]$ increasing as $(\zeta_0+\delta \zeta)/\zeta_0$ increases, 
and decreasing as $(\zeta_0+\delta \zeta)/\zeta_0$ does. 
Here the recombination rate
becomes larger than the ionisation rate immediately after the peak of
the X--ray flare, causing the ionisation fraction to decrease
as the flare intensity diminishes. The ionisation fraction returns
to the value obtained for a steady X--ray flux with ionisation rate
$\zeta_0$ shortly after the flare finishes, and remains at this value
until the onset of the next flare. As was the case with the
Oppenheimer \& Dalgarno model, this fall of the ionisation fraction
to the base value occurs because the recombination time is shorter
than the time period between X--ray flares. \\
\indent
For regions beyond $R \ge 2 \ \rm AU$ we evolved the chemistry for
times beyond $t = 10, 000 \ \rm yr$ until limit cycle behaviour
was obtained. These limit cycles were very similar to those
shown in the right panel of figure~\ref{figure4}, having similar
amplitudes for the case without magnesium, and very small amplitudes
when magnesium was included with $x_{\rm Mg}=10^{-8}$.


\subsection{Results with varying plasma temperature $k_B T$} 
\label{sec3.3}
In the preceding sections we discussed the effects of changing
the X--ray luminosity during a flare on the ionisation structure
of a disk model, while keeping the plasma temperature constant.
Observations, however, indicate that the plasma temperature $k_B T$
increases during flaring activity, resulting in the hardening
of the X--ray spectrum and an increased ability of X--ray photons to penetrate
into the disk (Wolk et al. 2005). Modelling of X--ray 
sources often requires a two--phase
emission model (Mewe, Kaastra \& Liedahl 1995), corresponding to two different
temperatures for the cooler and hotter plasma components of the
corona. Observations indicate that the cooler component remains
essentially unaffected during a flare, while the hotter component
increases significantly. In our model we assume that $k_{\rm B} T_{\rm cool}=3$
keV and $k_{\rm B} T_{\rm hot}=7$ keV, in close agreement with observations.

The requirement that we temporally resolve the X--ray flares
while solving the kinetic equations for the chemistry
means that a maximum time--step of 1 hour was adopted. This causes
the calculations to become very expensive computationally, such that for
the UMIST model \texttt{model3} a calculation lasting for 10,000 yr
requires a run time of approximately three weeks
for each grid--point of the disk model.
As a consequence we have been forced to only consider the 
Oppenheimer \& Dalgarno model \texttt{model1} in the following sections.
We simply note at this point that \texttt{model1} tends to generate
thinner dead--zones than \texttt{model3}, such that the results
described below represent an optimistic view of which fraction of the disk
can sustain MHD turbulence. We note further, however, that similar trends in the
results are obtained with \texttt{model1} and \texttt{model3} when varying
physical parameters, such that \texttt{model1} gives a reasonable
picture of how the ionisation fraction obtained using the more complex 
\texttt{model3} responds to changes in ionisation rate etc.

We begin by presenting the results of a model in which we 
consider the effects of increasing the plasma temperature during an
X--ray flare, keeping the X--ray luminosity $L_X$ constant.
This allows us to isolate the effects of the hardening of the X--ray spectrum
due to the rising plasma temperature.
We then consider a model for which both the plasma temperature
and the X--ray luminosity increase during flaring activity. \\

\noindent {\bf Snapshot at 10,000 yr with $L_X$ constant and $k_B T$ varying} \\
The results for this calculation are shown in the left panel of 
figure~\ref{figure6} which shows the column density of the whole disk
plotted as a function of radius using the solid line. The dashed lines
show the column density of the active zone for two calculations
that assumed constant plasma temperatures of $k_B T=3$ keV and
$k_B T=7$ keV, respectively, and the dotted line corresponds to 
the model whose plasma temperature rose to $k_B T=7$ keV during
flares only without an accompanying increase in $L_{\rm X}$ above 
$10^{30}$ erg s$^{-1}$.
Note that these figures refer to
\texttt{model1} with no heavy metal included.

\begin{figure}[t]
\psfig{silent=,figure=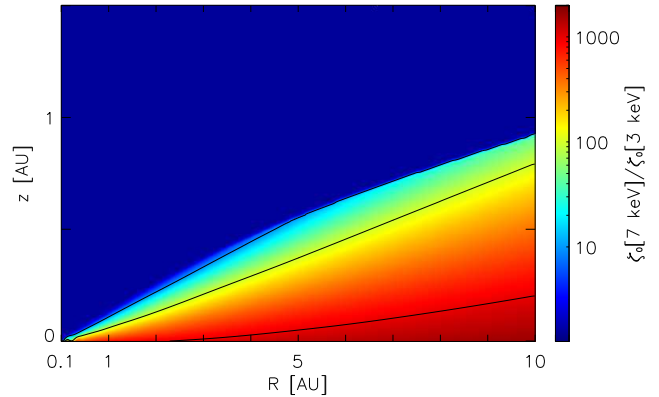,width=9.0cm}
\caption[]{The ratio of the effective X-ray ionisation rates
$\zeta_{0 ,7 \ {\rm keV}}/\zeta_{0 ,3 \ {\rm keV}}$.
$\zeta_{0 ,3 \ {\rm keV}}$ and $\zeta_{0 ,7 \ {\rm keV}}$
refer to local ionisation rates  for plasma temperatures
$k_{\rm B}^{}T = 3 \ \rm keV$ and $k_{\rm B}^{}T = 7 \ \rm keV$,
respectively. The disk parameter are $\alpha = 10^{-2}$ and ${\dot M} = 10^{-7}$
M$_{\odot}$yr$^{-1}$. The contour lines refer to values of
$\zeta_{0 ,7 \ {\rm keV}}/\zeta_{0 ,3 \ {\rm keV}}$: $10, 10^2$, and $10^3$.
\label{figure7}}
\end{figure}

Figure~\ref{figure7} shows contours of the relative ionisation rate
when the plasma temperature is raised from $k_B T=3$ to 7 keV.
The change in $\zeta$ is dramatic in the deeper, more shielded regions
simply because of increased penetration induced
by the hardening of the X--ray spectrum. Near the disk surface the change
is only slight, but in the interior the local ionisation rate can
increase by more than a factor of 1000.

The upper dashed line in the left panel of figure~\ref{figure6} shows that
having a constantly increased value of $k_B T=7$ keV significantly
reduces the depth of the dead-zone throughout the disk. The dotted line,
however, shows that at a point in time midway between two peaks in the
plasma temperature, the dead--zone is unaffected interior to $R <1.2$ AU
because of the fast recombination time there. Beyond 
$R>2$ AU the dead--zone is a little thinner where the recombination time
is longer than the period between increases in plasma temperature,
corresponding closely to that which would be obtained by exposure
to the time averaged value of the time dependent ionisation rate. \\

\noindent {\bf Results with $L_X$ and $k_B T$ varying} \\
A snapshot of the results obtained after 10,000 yr 
when the X--ray luminosity {\em and}
the plasma temperature increase during a flare is shown in 
the right panel of
figure~\ref{figure6}. The solid line gives the column density of the
whole disk, and the dashed line gives the column density of the active zone
obtained when the X--ray luminosity is constant and takes its
base value $L_{\rm X}^0$. The disk becomes active everywhere if we take a 
constant ionisation rate with $L_{\rm X}=100 L_{\rm X}^0$ and $k_{\rm B} T=7$ keV.
The dotted line corresponds to
the model for which the plasma temperature and X--ray luminosity
increase during a flare. Note that these models are \texttt{model1}
in the absence of heavy metals, and the snapshot is taken at
a time that is midway between two X--ray flares.
In the inner regions $R<1.2$ AU the dead--zone size is again
unaffected by the flares between outbursts because of the rapid
recombination time there.
Beyond $R>2$ AU, however, the dead--zone depth has been decreased substantially
such that the column density of the active zone includes approximately
80 \% of the matter in the region between $R=$ 3 - 7 AU.
Although not plotted in figure~\ref{figure6}, the addition of heavy metals
with $x_{\rm M}=10^{-12}$  
causes the dead zone to disappear completely beyond $R>2$ AU, although a
dead--zone ranging between $0.4 < R < 2$ AU remains.

\begin{figure*}[t]%
\hbox{%
\psfig{silent=,figure=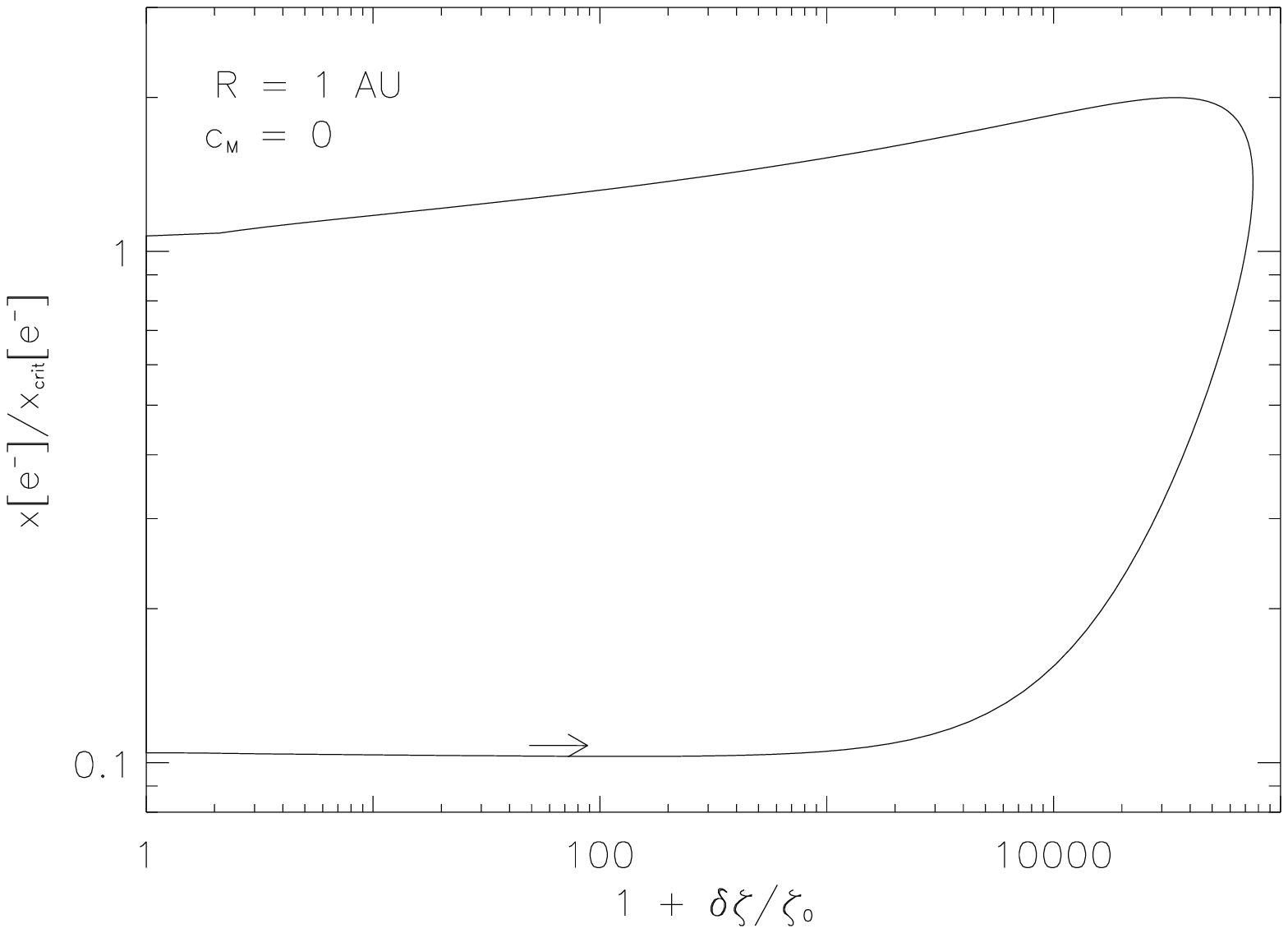,width=9.0cm} \hspace*{-.6cm}
\psfig{silent=,figure=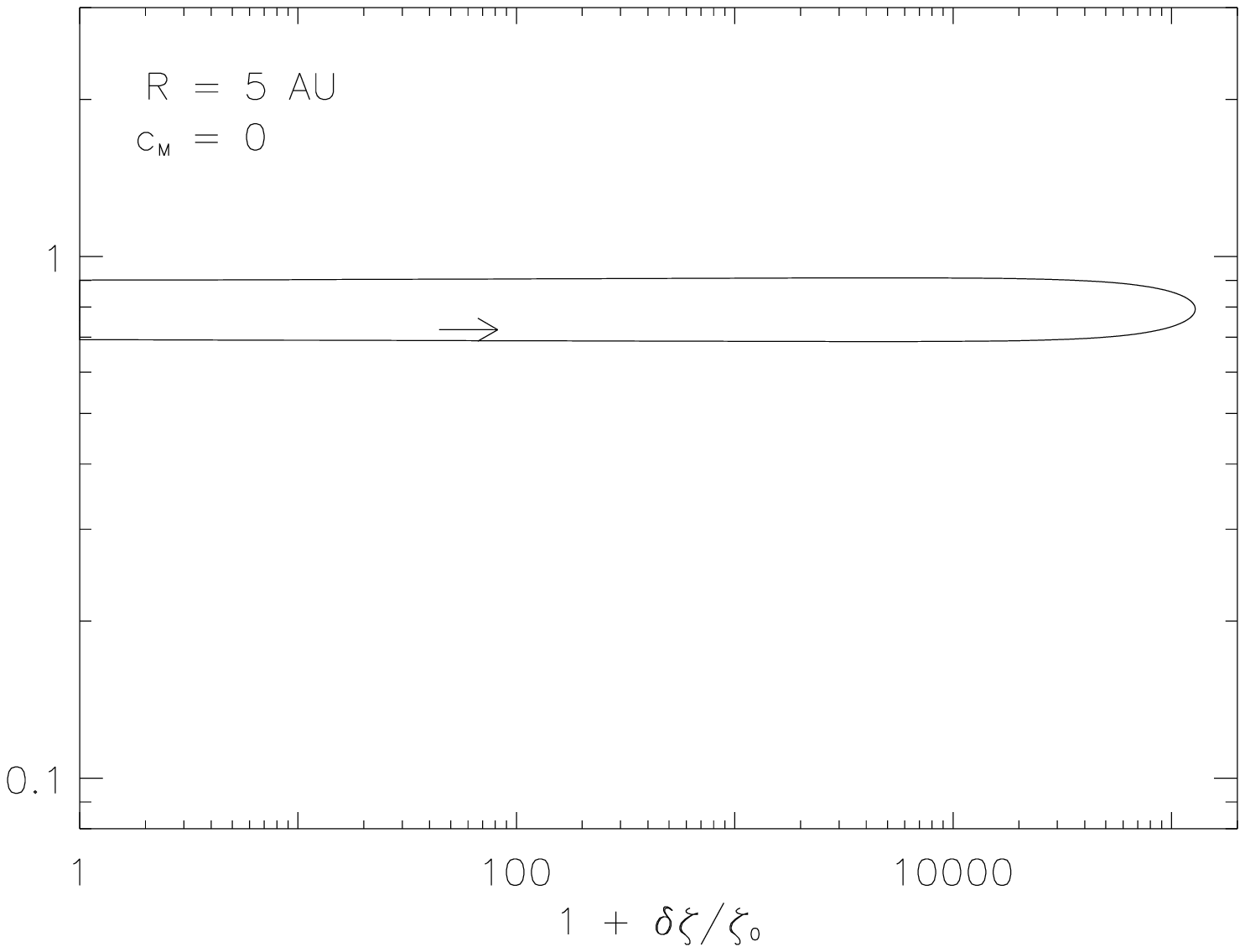,width=9.0cm}}
\caption{\texttt{model1} - The change in the ionisation fraction
$x[\rm e_{}^{-}]/x_{\rm crit}[ e_{}^{-}]$ is plotted against the
relative change in
the ionisation rate $(\zeta_0 + \delta \zeta)/\zeta_0$ at
different radial positions
$R$ with $z/R = 0$ (midplane). $x_{\rm crit}[ e_{}^{-}]$ denotes
the ionisation fraction
at the transition layer. Each graph describes a limit cycle since the orbit
is closed (the arrow points forward in time). The perturbation
$\delta \zeta/\zeta_0$ in
the ionisation rate is
based on the assumption that both the X-ray peak luminosity
$L_{\rm X}^{P} = 100 \times L_{\rm X}^{0}$ and the temperature of
the hot plasma
component $k_{\rm  B}^{}T_{\rm hot}^{} = 7 \ \rm keV$ at the flare peaks.
\label{figure8}}
\end{figure*}

We now consider the time dependent evolution after 10,000 yr has elapsed.
The evolution of the ionisation fraction at the disk
midplane (and not at the transition zone as in previous similar figures)
as a function of the perturbed ionisation rate is shown in 
figure~\ref{figure8} at radii $R=1$ AU and $R=5$ AU for
\texttt{model1} in the absence of heavy metals. The first thing to note is
the large perturbation to the ionisation rate experienced
at the disk midplane, as illustrated by the range of values for
$(\zeta_0+\delta \zeta)/\zeta_0$. 
Second, the ionisation fraction, relative to the
critical value $x_{\rm crit} [{\rm e}^{-}]$, at the midplane located at
$R=1$ rises from 0.1 (implying the region is dead) to a value of $\approx 2$
(implying that the region is active). Thus, the inner regions of the disk
show the limit cycle behaviour, and oscillate 
from having a dead--zone near the midplane to having an
active zone there which lasts for a duration slightly longer than half
the flare (i.e. about 12 hours). This region remains formally ``dead'' for
the other six and a half days during the X--ray flaring cycle.\\
\indent
Although at first sight this would appear to indicate that the 
stress associated with MHD turbulence in a protoplanetary disk
would be episodic, leading
to episodic mass accretion, it is important to bear in mind that
the local growth rate for the MRI is approximately the local dynamical
time (Balbus \& Hawley 1991),
which is longer than 12 hours in the regions beyond 0.5 AU.
It is therefore unclear whether such a region of the disk could
maintain fully developed turbulence and an associated dynamo
when perturbed by X--ray flares of short duration. This issue is discussed 
below.\\
\indent
The right panel of figure~\ref{figure8} shows the evolution of
$x[{\rm e}^{-}]/x_{\rm crit} [{\rm e}^{-}]$ versus the perturbation to
the local ionisation rate $(\zeta_0+\delta \zeta)/\zeta_0$ at the disk midplane
at $R=5$ AU. Here it is clear that limit cycle behaviour is again
obtained, with the ionisation fraction showing a closed orbit.
In this part of the disk, there remains a thin dead--zone throughout the
cycle when heavy metals are not present. The addition of heavy metals
with abundance $x_{\rm Mg}=10^{-12}$ renders the disk fully active in
this region for the whole duration of the X--ray cycle when both the 
luminosity and plasma temperature vary.

\section{Discussion}
\label{sec4}
In this paper we have calculated the response of the ionisation
fraction in a standard $\alpha$-disk model to the presence of
X--ray flares originating in the corona of the central T Tauri star.
For comparison purposes we have examined how the disk responds to 
constantly acting X--ray induced ionisation rates $\zeta_0$
(the base ionisation rate), $\zeta_{\rm eff}$ (the time averaged ionisation
rate), and $100 \zeta_0$ (which is the peak ionisation rate during a flare).
We have considered cases where the plasma temperature retains a
constant value of $k_{\rm B} T=3$ keV during the flare cycle,
and cases where the plasma temperature increases from 3 keV to 7 keV at
the peak of a flare.
Broadly speaking, the behaviour of
the dead--zones in our disk model can 
be divided into three basic
regions when subject to X--ray flares: \\
({\it i}). An inner region $R< 0.5$ AU where the disk is entirely active due to
thermal ionisation of potassium throughout the calculations. \\
({\it ii}). A central region $0.5 < R < 2$ AU where the disk exhibits
a significant dead--zone between outbursts, but whose dead--zone
can be significantly diminished or be removed altogether near the
peak of an outburst. This behaviour occurs because of the short
recombination time here, even when heavy metals are included in the gas
phase due to the recombination being dominated by molecular ions.
This causes the ionisation fraction 
to remain close to being in--phase with the ionisation rate during and
between flares. Our calculations indicate that allowing the plasma
temperature to increase to $k_{\rm B} T=7$ keV
during X--ray flares can cause the 
magnetic Reynolds number to exceed 100 for about 12 hours after the peak of 
the flare, such that the dead--zone here is formally removed for that time. \\
({\it iii}). An outer region $R>2$ AU where the dead--zone size and depth
does not
change significantly during the flaring cycle. Here the dead--zone size 
corresponds closely to that obtained by the time--averaged 
X--ray flux, because the
recombination time exceeds the time period of the X--ray outburst cycle.
The depth of the dead--zone size in this region is significantly diminshed
if heavy metals are present in the gas phase {\em or}
the plasma temperature increases during flares. The dead--zone may disappear 
altogether if metals are present {\em and} the plasma temperature
increases during flares. In the absence of heavy metals and
an increasing plasma temperature, the time averaged
ionisation rate during a flare cycle $\approx 5.8 \zeta_0$,
and a significant dead--zone remains.

Given this disk structure consisting of an inner region without a dead--zone,
a central region whose dead--zone can change significantly or be removed
during parts of the X--ray flaring cycle, and an outer region which may
maintain a thin dead--zone of almost constant depth, it is interesting to
speculate how MHD turbulence in the system will evolve in time.
Clearly the innermost regions can sustain MHD turbulence continuously,
as can the magnetically active component of the outer disk.
The behaviour of the central region whose magnetic Reynolds number
varies significantly throughout the X--ray flaring cycle is less clear.
In this region the dead--zone remains deep for six and a half days
and is formally reduced in size or removed for only about 12 hours.

In the ideal MHD limit the fastest growing mode of the MRI
has wave number $k_{\rm max}$
defined by $k_{\rm max} v_A \approx \Omega$, where $v_A$ is the Alfv\'en speed
and $\Omega$ is the local Keplerian angular velocity.
The growth rate associated with this mode is $\approx \Omega$ (actually closer
to $0.75 \Omega$ [Balbus \& Hawley 1991]).
Ohmic resistivity strongly affects the growth of linear modes with wave number
$k$ when the associated diffusion rate $k^2 \eta$ exceeds the linear growth rate.
Comparing these rates for the fastest growing mode provides a condition
for the effects of resistivity to dominate over the growth of the mode
due to the MRI:
\begin{equation}
k_{\rm max}^2 \eta > k_{\rm max} v_A
\label{rates1}
\end{equation}
which yields the condition
\begin{equation}
\frac{v_A^2}{\eta \Omega} < 1.
\label{rates2}
\end{equation}
Equation~(\ref{rates2}) is often used to provide an alternative definition
of the magnetic Reynolds number $Re_{\rm m}'=v_A^2/(\eta \Omega)$
(e.g. Sano, Inutsuka \& Miyama 1998) which is related to the definition
used in this paper by
\begin{equation}
Re_{\rm m} = \frac{\beta}{2} Re_{\rm m}'.
\label{Re_nums}
\end{equation}
where $\beta = 2 c_s^2/v_a^2 = P_{\rm gas}/P_{\rm mag}$.
$P_{\rm gas}$ and $P_{\rm mag}$ are the gas and magnetic pressure.
Thus a value of $Re_{\rm m}=100$ corresponds to a value of
$Re_{\rm m}'=1$ when $\beta=200$, such that resistivity
dominates over the growth of the fastest growing mode for
values of $Re_{\rm m}$ and $Re_{\rm m}'$ smaller than these.

For the purpose of illustration, let us consider a situation where $\beta=200$
in the central regions of our disk near the location $R=0.5$ AU,
where the inverse of the Keplerian angular velocity $\Omega^{-1}=20.5$ days.
An X--ray flare arises, raising the value of $Re_{\rm m}> 100$ for
12 hours, allowing exponential growth of the fastest growing mode
for this time such that an amplification of $\approx 2.5 \%$ occurs.
Given our definition of $Re_{\rm m}$ it is clear that
any field amplification that occurred while $Re_{\rm m} > 100$
is diffused away within the next 12 hours after $Re_{\rm m}$ drops below 100,
and the growth of this mode is unable to amplify the field or
drive the disk toward a turbulent
state. The situation can in principle be different for longer wavelength
modes. Consider the longest wavelength mode that can fit within 
the disk vertical extent with wavelength $\lambda=2 H$, and associated
wavenumber $k_{\min}=\pi/H$. The wave number associated with the fastest
growing mode has $k_{\rm max}=10/H$ if $\beta=200$, such
that $k_{\rm max}/k_{\rm min} \approx 3$. The diffusion time associated with
the longest wavelength mode is an order of magnitude longer than for
the fastest growing mode (whereas the growth rate is approximately half of
the maximum value). Thus
any field amplification that occurs during the 12 hours when
$Re_{\rm m}>100$ is diffused away over the next $\approx 5$ days.
This suggests that an X--ray flaring cycle with a periodicity of
less than five days could in principle lead to gradual field
amplification over successive cycles. The question of whether
fully developed turbulence can arise in such a scenario is unclear,
and can only be addressed by means of non linear simulations
that explicitly account for periodic rises in the ionisation rates 
due to X--ray flaring and electron recombination.

The calculations presented in this paper consider only gas--phase chemistry,
and ignore the effects of dust. As such they are relevant to a 
stage in protoplanetary disk evolution when substantial grain growth 
has occurred and a dense dust layer has settled near the midplane.
It is well known, however, that small dust grains are able to
sweep up free electrons and substantially reduce the ionisation
fraction (e.g Sano et al. 2000; Ilgner \& Nelson 2006a), and
the assumption of gas--phase chemistry is only really valid
when the abundance of dust grains has been depleted by a factor
of between $10^{-4}$ -- $10^{-8}$ below the canonical concentration of
$10^{-12}$ (Ilgner \& Nelson 2006a).
An open question that we have not addressed 
is what happens to species that are adsorbed onto the surfaces
of small grains during X--ray flares, and in
particular what happens to the adsorbed electrons. Najita et al. (2001)
considered the effect of nonthermal desorption of grain mantles
due to X--rays, but did not include the desorption of
electrons into the gas phase. We speculate that if grain mantles
and electrons are  desorbed into the gas phase
during X--ray flares, and are adsorbed back onto the grains
during the time between flares, then the behaviour of the dead--zones
throughout the disk will be similar to that already
observed in the central regions between $0.5 < R < 2$ AU.
This is because electrons adsorb onto grains very rapidly
throughout the disk, such that the time scale of which 
electrons will be removed from the gas phase is shorter than the
time between flares. An analysis of this issue will be
presented in a future publication.

\section{Summary}
\label{sec5}
We have considered the effect of X--ray flares on the ionisation 
fraction and dead--zone structure in protoplanetary disks. 
These flares can have a significant effect on dead--zones if much of 
the submicron sized dust in the disk has undergone grain growth
and settled toward the midplane, if the plasma temperature increases
significantly during X--ray flares, and/or trace quantities of heavy metals 
(magnesium) are present in the gas phase.\\
\indent
Questions remain, however, about the disk response in regions where 
the dead--zone is removed during an X--ray flare, but reappears 
during the low state. In particular it is not clear that MHD turbulence 
can be generated there as the growth time of the MRI is longer than 
the duration of the flares. This issue needs to be addressed using non 
linear simulations that take account of the time dependent ionisation 
rates induced by X--ray flares.


\begin{acknowledgements}
\noindent
This research was supported by the European Community's Research
Training Networks Programme under contract HPRN-CT-2002-00308,
"PLANETS". The calculations presented here were performed using the
QMUL HPC facility purchased under the SRIF initiative. We wish to
thank Eric Feigelson for information provided concerning X--ray flares
during the early stages of this project.
\end{acknowledgements}


\end{document}